\documentclass{amsart}
\usepackage{epsfig}
\usepackage{times}
\usepackage{amsmath}
\usepackage{amssymb}
\usepackage{xspace}
\usepackage{algorithmicx}
\usepackage{algpseudocode}
\usepackage{pb-diagram}
\usepackage{graphicx}
\newcommand {\mm}[1] {\ifmmode{#1}\else{\mbox{\(#1\)}}\fi}

\newsavebox{\smallProofsym}                            % smallproofsymbol
\savebox{\smallProofsym}

\makeatletter
\long\def\@makecaption#1#2{%
  \vskip\abovecaptionskip
  \sbox\@tempboxa{\small #1: #2}%
  \ifdim \wd\@tempboxa >\hsize
    \small #1: #2\par
  \else
    \global \@minipagefalse
    \hb@xt@\hsize{\hfil\box\@tempboxa\hfil}%
  \fi
  \vskip\belowcaptionskip}
\makeatother

\newcommand{\R}        {\mm{{\mathbb R}}}
\newcommand{\Z}        {\mm{{\mathbb Z}}}
\newcommand{\X}        {\mm{{\mathbb X}}}
\newcommand{\U}        {\mm{{\mathbb U}}}
\newcommand{\Y}        {\mm{{\mathbb Y}}}

\newtheorem{theorem}{Theorem}[section]

\begin{document}

\title{Multi-Scale Local Shape Analysis and Feature Selection
in Machine Learning Applications}

\author{Paul Bendich}
\address{Department of Mathematics, Duke University}
\email{bendich@math.duke.edu}

\author{Ellen Gasparovic }
\address{Department of Mathematics, Duke University}
\email{ellen@math.duke.edu}

\author{John Harer}
\address{Department of Mathematics, Duke University}
\email{harer@math.duke.edu}

\author{Rauf Izmailov}
\address{Applied Communication Sciences}
\email{rizmailov@appcomsci.com}

\author{Linda Ness }
\address{Applied Communication Sciences}
\email{lness@appcomsci.com}

\begin{abstract}

We introduce a method called \emph{multi-scale local shape analysis} for
extracting features that describe the local structure
of points within a dataset. The method uses both 
geometric and topological features at multiple levels of granularity
to capture diverse types of local information for subsequent machine
learning algorithms operating on the dataset. Using synthetic and
real dataset examples, we demonstrate significant performance
improvement of classification algorithms constructed for these
datasets with correspondingly augmented features.

\end{abstract}

\maketitle

%%%%%%%%%%%%%%%%%%%%%%%%%%%%%%%%%%%%%%%%%%%%%%%%%%%%%%%%%%%%%%%%%%%%%%%%%%
\section{Introduction}
\label{sec:intro}

The goal of this paper is to introduce a preliminary version of what
we call \emph{multi-scale local shape analysis} (MLSA), a
method for extracting features of a dataset that describe the local
structure, both manifold and singular, of points within the dataset.
MLSA is a mixture of \emph{multi-scale local principal component
analysis} (MLPCA) and \emph{persistent local homology} (PLH). In
this paper, we will describe both of these techniques and our merger
of them, and we will demonstrate the potential of MLSA on two
synthetic datasets and one real one.

The potential of these methods and their merger is investigated in
the context of one of the typical applications for data analytics:
the classification problem for multi-dimensional datasets. Thus
the relevance of the developed techniques is assessed as the quality
of the resulting classification decision rule, measured by the
expected test misclassification error, its sensitivity and
specificity (false positive and false negative error rates).

The quality of the solution of the classification problem
significantly depends on the choice of the features -- specifically,
on (1) extraction of new features that can contain additional
relevant information for the given problem, (2) pre-processing the
features in a way that makes them feasible for scalable and robust
computations, and (3) removing features that have little relevancy
for the problem. While there are well-known mechanisms of removing
features (i.e., feature selection, see \cite{LiuMotoda},
\cite{Breiman}), the problem of constructing or adding features
\cite{Domingos} is much more challenging (see \cite{GuyonElisseeff},
\cite{Torkkola}), since it often relies on domain expertise, which
is difficult to automate.

That is why, besides domain expertise, numerous geometrical
approaches for feature extraction have been employed to reduce the
misclassification error rate
of the decision rule (e.g., kernel PCA \cite{Scholkopf}, mutual
information \cite{Torkkola}, manifold learning \cite{Jones08},
\cite{CoifmanLafon}, image-derived features \cite{Bow},
\cite{Sonka}). The added benefit of these methods stems from the fact that
geometric methods can expose additional relevant information about
shapes that are hidden in the original data. The methods
outlined in this paper capture both geometric (MLPCA) and
topological (PLH) structures of datasets, thus exposing the
corresponding structures to machine learning tools and boosting
their performance.

%\paragraph{MLPCA}

{\bf MLPCA.} Principal Component Analysis (PCA) is a standard
technique that takes a point cloud $\X \subseteq \R^D$ as input, and
returns information about the directions of data variation and their
relative importance. A standard output of PCA is an integer $k$
and a projection of the data onto the $k$-dimensional linear
subspace of $\R^D$ spanned by the $k$ most important directions. In
this case, it is reasonable to say that the ``intrinsic dimension''
of $\X$ is $k$ in the sense that if one ``zooms in'' at a point in
$\X$, it will look like a $k$-flat. Of course, for some datasets the
intrinsic dimension varies as we move around the data (think of a
plane pierced by a line in $\R^3$). For almost any dataset, the
notion of ``intrinsic dimension'' depends on how much one wants to
zoom in. For example, data sampled from a thin strip in $\R^2$ will
look either one- or two-dimensional, depending on the notion of
local scale.

%\paragraph{PLH}

{\bf PLH.} To complicate matters further, there are datasets $\X$ and
points $z$ for which the statement ``Locally at $z$, $\X$ looks like
a $k$-flat'' is simply not true, for any value of $k$. For example,
consider a dense sample from the intersection of two planes in $\R^3$.
If $z$ is any point very near to the line of intersection, then MLPCA at almost any radius will return
a $3$-flat, which is not a reasonable answer. The key issue here is that
proximity to singularities complicates the notion of local dimension.
It is here where PLH proves useful.

The concept of PLH is built off of a traditional algebro-topological
notion called \emph{local homology groups}; see, for example,
Chapter $35$ in \cite{Munkres2}. These groups are meant to assess
the ``local'' structure of a point $z$ within a topological space
$\X$. Among their nice properties is the fact that they are the same for
every point $z$ precisely when $\X$ is a topological manifold.
When $\X$ is not, these groups differ as
we move $z$ around, and in fact provide a great deal of information
about the local singularity structure at each non-manifold point.

On the other hand, the concept of ``local'' is a tenuous one in the
noisy point-cloud context, where what is meant by local depends
entirely on an often impossible-to-choose scale parameter. This
issue was addressed in \cite{BCE07}, where a tweakable radius
parameter $R$ was added to the definition. The version of local
homology that we will use in this paper differs slightly from that
in \cite{BCE07}, but we feel it is simpler both for exposition and
for computational purposes.

{\bf Related work.} To the best of our knowledge, this is the first attempt to use PLH
in the construction of features for classification problems.
PLH has been used before \cite{DFW14} in the context of dimension reduction.
However, the goal there is to use PLH to detect the dimension of the manifold
which, under the assumptions of that paper, underlies the given dataset.
Our goal is quite different: to augment a more standard dimension-detection
method with PLH in order to understand features that may arise by dropping
the underlying-manifold assumption.

Another paper \cite{bendich2012stratlearn} uses PLH, and also a more complicated
mapping construction to transfer PLH information from one point to another, in an
effort to learn the underlying stratified structure of a space from a point sample of
it.
There is also some work \cite{Sayan2014grassman} which learns the stratified structure
of a union of flats via Grassmannian methods.

Finally, a recent paper \cite{AhmedFasy2014} uses PLH in the construction of a novel distance between
different road map reconstructions.

%\paragraph{Outline}

{\bf Outline.} The structure of this paper is as follows. In Section
\ref{sec:MLPCA}, we briefly review the ideas behind MLPCA, followed
by a more in-depth description of PLH in Section \ref{sec:PLH}.
Then, in Section \ref{sec:results}, we combine the two techniques into MLSA and demonstrate 
its utility in machine learning experiments involving three sample datasets, two synthetic and one real.
The results of these experiments are summarized in the tables of Section \ref{sec:ST}.
We conclude with some discussion in Section \ref{sec:disc}.
%%%%%%%%%%%%%%%%%%%%%%%%%%%%%%%%%%%%%%%%%%%%%%%%%%%
%%%%%%%%%%%%%%%%%%%%%%%%%%%%%%%%%%%%%%%%%%%%%%%%%%%

\section{Multi-Scale Local Principal Component Analysis}
\label{sec:MLPCA}

\begin{figure}
%\vspace{.3in}
\centerline{\fbox{\includegraphics[scale=0.25]{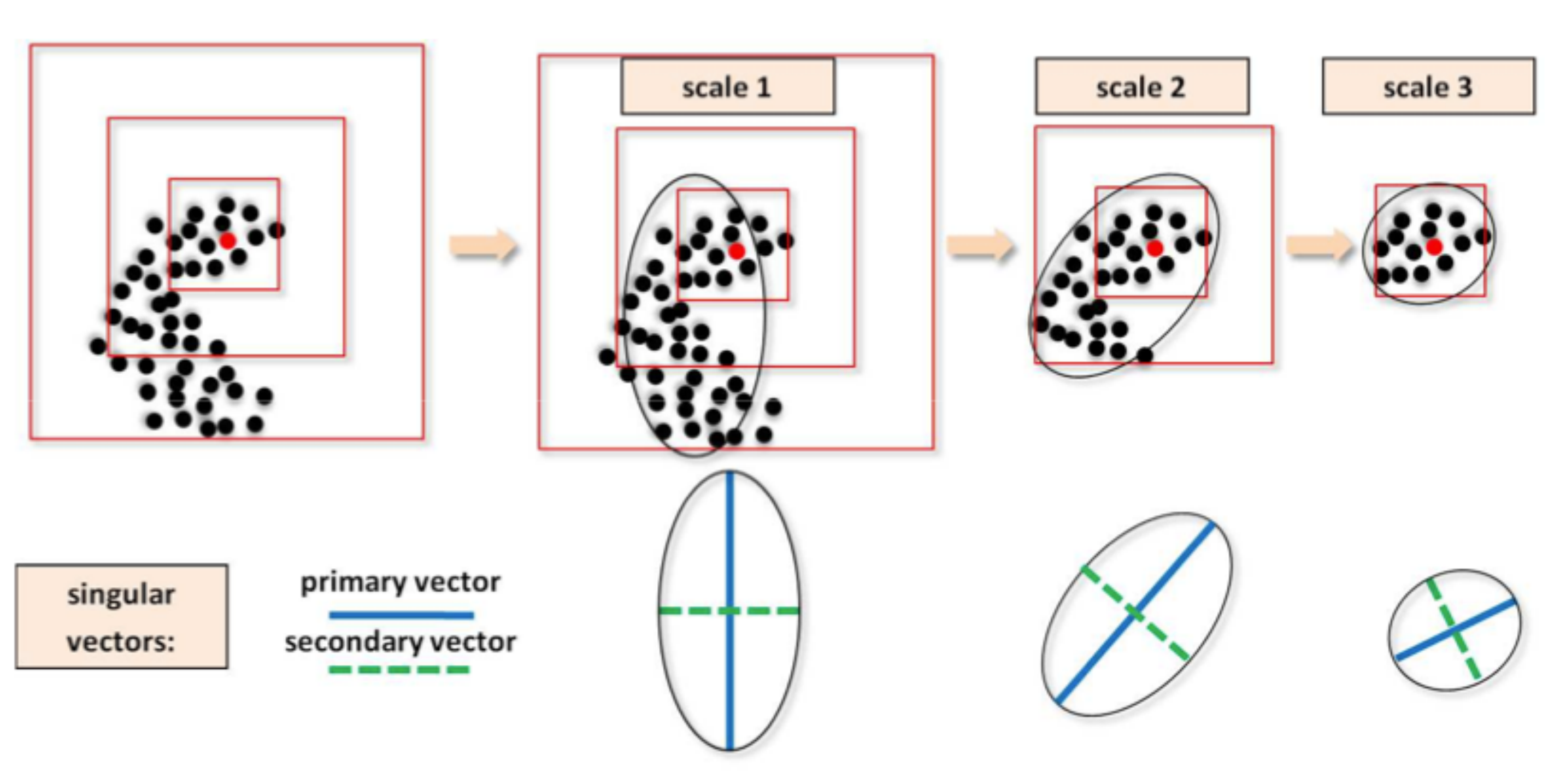}}}
%\vspace{.3in}
\caption{Multi-scale local principal component
analysis (MLPCA).} \label{fig:MLPCA}
\end{figure}

In \emph{multi-scale local principal component analysis} (MLPCA)
\cite{Le03}, one takes a point cloud $\X$, a particular point $z$
and a radius $R$, and one computes PCA on the sub-cloud of points
within the Euclidean $R$-ball around $z$. This process is then
repeated for multiple radii and at many points to get a general
profile of how the dataset looks at different locations in different
scales (Figure~\ref{fig:MLPCA}). These
multi-scale features may then be used in the overall data analysis
along with the original features (coordinates of the points). 

This approach belongs to a growing
family of multi-scale methods exploring and leveraging natural
differences in information availability and its relevance on
different scales of the dataset in question. Multi-scale methods
tend to reveal more accurate information about the structure of the
dataset than global (single scale) techniques.

The eigenvalues and eigenvectors of the covariance matrix
provide a core set of features which capture geometric
information about the dataset. The first $k$ eigenvectors of
the covariance matrix define a $k$-dimensional hyperplane
(through the center of mass) minimizing $\sum_i({\rm distance}
(x_i,P))^2$, where the infimum is taken over all $k$-planes $P$.
This geometric information  can be
significantly enriched by computing the eigenvalues and
eigenvectors for a set of multi-scale neighborhoods of
points. Bassu et.
al.~\cite{BIMNS}  exploited multi-scale local PCA features to define
a support vector machine (SVM) decision rule that distinguished pointwise two unknown
empirical measures (ground and vegetation) on the same domain of
LIDAR-generated surface images. The same authors later applied this
technique for classification of various types of satellite images of
vessels~\cite{BIMNS2}. These two image analysis experiments
demonstrate that centralized multi-scale PCA features on data
sets in $\mathbb{R}^n$ provide necessary and sufficient conditions for datasets
to have certain properties.

%%%%%%%%%%%%%%%%%%%%%%%%%%%%%%%%%%%%%%%%%%%%%%%%%%%

\section{Persistent Local Homology}
\label{sec:PLH}

This section contains a formal description of PLH features, which will be
used in the next section to augment MLPCA features
in several example applications.
We assume that the reader understands homology groups, and we give only the
briefest of reviews of persistent homology.
For a good reference on the former, see \cite{Munkres2}; for the latter,
see \cite{Edelsbrunner2010} or \cite{Chazal2009b}.
All homology groups need to be computed over a field for the definition of
persistence to make sense, and that field
is usually $\Z / 2\Z$ for computational reasons.

\subsection{Persistent Homology}

Suppose that $\X$ is a topological space equipped with a real-valued function $f$.
For each real number $\alpha$, we define the \emph{threshold set}
$\X_{\alpha} = \{x \in \X \mid f(x) \leq \alpha\}$.
Note that increasing $\alpha$ from negative to positive infinity provides
a \emph{filtration} of $\X$ by
these threshold sets.
For each non-negative integer $k$, the \emph{persistence diagram} $Dgm_k(f)$
summarizes the appearance and disappearance
of $k$-th homology during this filtration.

\begin{figure}[h]
%\vspace{.3in}
\centerline{\fbox{\includegraphics[scale=0.3]{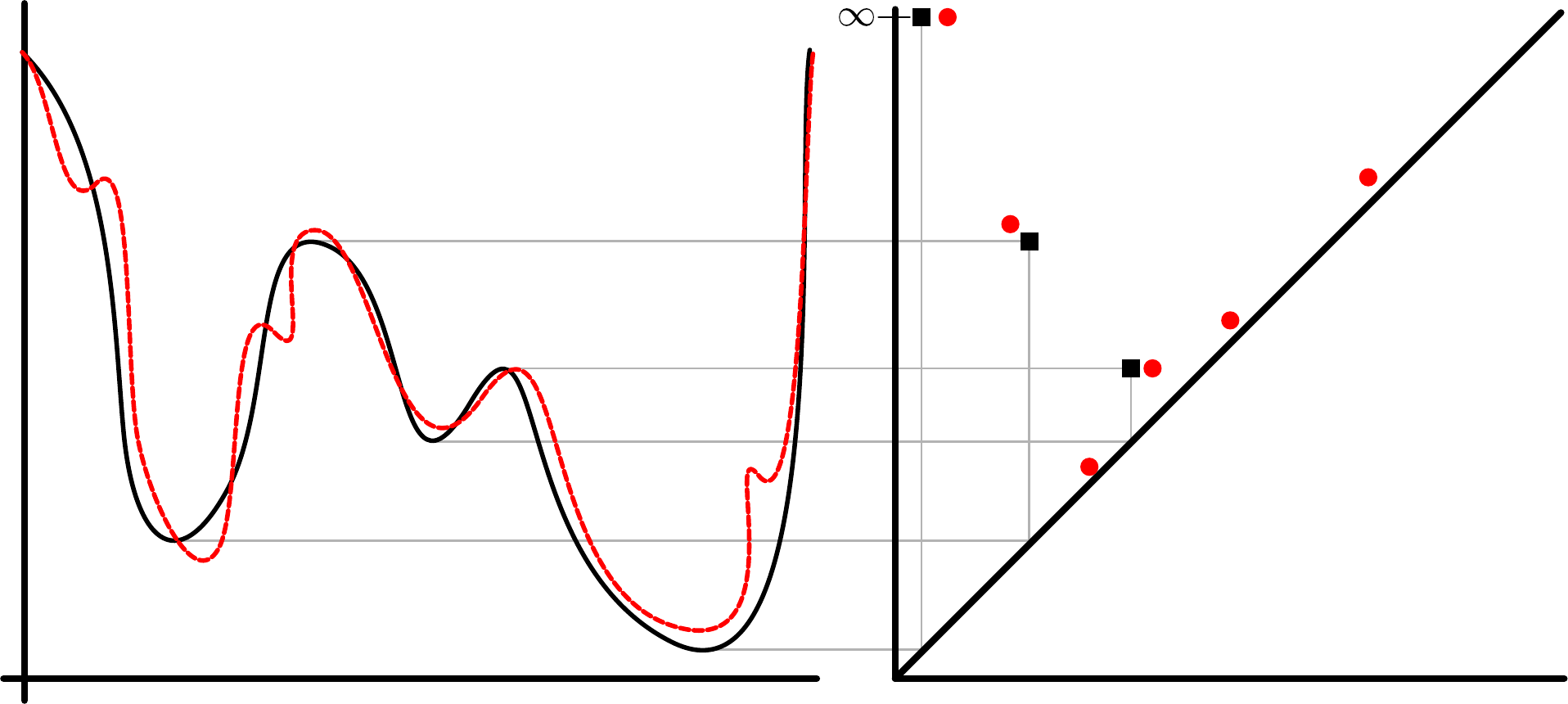}}}
%\vspace{.3in}
\caption{Left: the graphs of functions $f$ (black) and $g$ (red). Right: the
persistence diagrams
$Dgm_0(f)$ (black) and $Dgm_0(g)$ (red).}
\label{fig:pF}
\end{figure}

For example, take $\X$ to be a closed interval and $f$ to be the
function whose graph is drawn in black on the left of
Figure~\ref{fig:pF}. Then the persistence diagram $Dgm_0(f)$, shown
as black squares on the right of the same figure, encodes the
appearance and subsequent merging of components under this
filtration of the interval.

{\bf Stability theorem.} It turns out that the persistence diagram
$Dgm_k(f)$ is robust to small changes in the input function. To make
this more precise, we properly define a persistence diagram to be a
multi-set of dots in the extended plane, with the extra condition
that there is a dot of infinite
 multiplicity at each point
along the major diagonal $y = x.$
The \emph{persistence} of a dot $u = (x,y)$ in a diagram is defined to be $y - x$,
its vertical distance
to the major diagonal.

Given a $p \in [1, \infty)$, the $p$-th \emph{Wasserstein distance}
between two diagrams $D$ and $D'$ is
then defined to be:
\begin{equation}
 W_p(D,D') = \left [ \inf_{\phi:D \to D'} \Sigma_{u \in D} ||u - \phi(u)||^p \right ] ^{\frac1p},
\label{eqn:BD}
\end{equation}
where the infimum is taken over all bijections $\phi$ between the diagrams; note
that such bijections always exist, due
to the infinite multiplicity dots along the major diagonal.
Letting $p$ tend to infinity results in the \emph{bottleneck distance}
$W_{\infty}$ between the diagrams.

These distances are computed \cite{Edelsbrunner2010} via constructing
a minimal-weight perfect matching on a weighted bipartite graph, which means that
the distances themselves are not useful as an efficient tool.
However, they are important for stating the stability properties of persistence diagrams,
as we now illustrate via an example.

Let $g$ be the function whose graph appears in red on the left of
Figure \ref{fig:pF}, with diagram
$Dgm_0(g)$ given by red circles on the right.
Then the optimal bijection between the two diagrams would be the one that matches
the three red dots along the diagonal to the black diagonal, and the
other red dots to their closest black square, with the bottleneck distance being
the longest distance any red dot has to move during this process.

Note that the two diagrams are quite close under this metric, as are the two functions
under the $L_{\infty}$-metric.
This is true in general:
\begin{theorem}[Diagram Stability Theorem]
 Let $f$ and $g$ be two tame functions on a compact space $\X$.
Then, for each non-negative integer $k$, we have:
$$W_{\infty}(Dgm_k(f), Dgm_k(g)) \leq ||f - g||_{\infty}.$$
\label{thm:DST}
\end{theorem}
\vskip -0.2in See \cite{CohenSteiner2007} and \cite{Chazal2009b} for
a more technical discussion of this theorem, and see
\cite{Cohen-Steiner2010} for similar theorems, with more assumptions required,
about $W_p$-stability.

\subsection{Local Homology}

We now describe local homology groups, before moving on to their persistent version.

\begin{figure}[h]
%\vspace{.3in}
\centerline{\fbox{\includegraphics[scale=0.25]{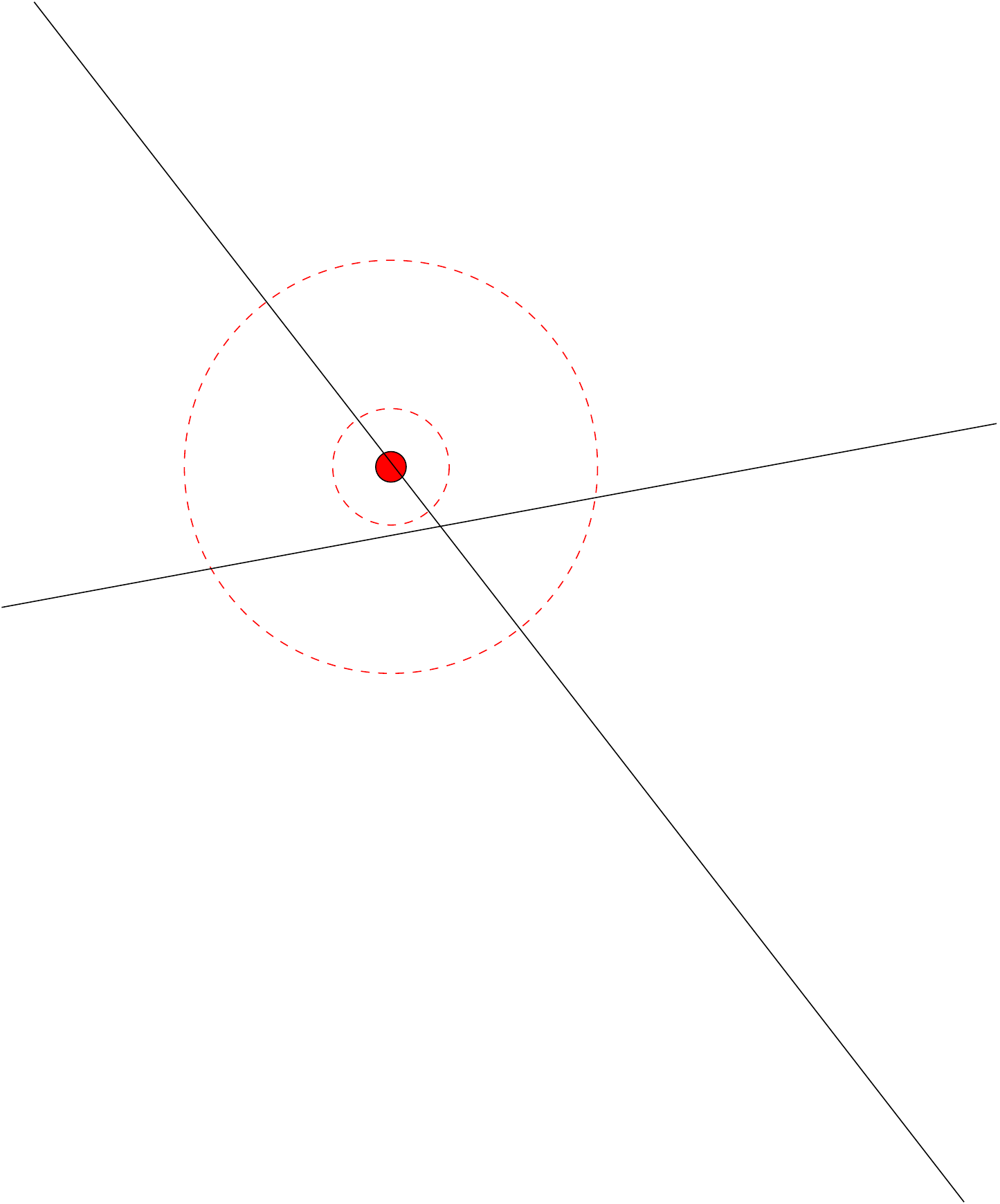}}}
%\vspace{.3in}
\caption{Within the smaller sphere, the red point looks like part of a
$1$-manifold; within the larger sphere, its local structure is more complicated.}
\label{fig:LH}
\end{figure}

Let $\Y$ be a topological space embedded in some $\R^D$, fix some point $z \in \R^D$,
a positive real number $R$, a non-negative integer $k$, and let
$S_R(z)$ denote the sphere of radius $R$ about $z$; that
is, $S_R(z) = \{y \in \R^D \mid ||y - z|| = R \}$.
The $k$-th \emph{local homology} group of $\Y$, with center point $z$ and
radius $R$, is defined to be
\begin{equation}
 LH_k(Y,z,R) = H_k(\Y \cap S_R(z)).
\label{eq:LH}
\end{equation}
For a simple example, take $\Y$ to be an infinite plane in $\R^3$, $z$ to be a
point on that plane, and $R$ to be any positive number.
Then $LH_k(Y,z,R)$ is rank one when $k = 1$ and is zero otherwise.
For a more complicated example, let $z$ be the red point in Figure \ref{fig:LH},
and let $r$ and $R$ be the radii of the smaller and large circles, respectively.
Then $LH_0(Y,z,r)$ and $LH_0(Y,z,R)$ are ranks two and four, respectively; both
groups are zero for all other $k$.

{\bf Instabilities.} As defined above, the local homology group
$LH_k(Y,z,R$) depends on three inputs, and it turns out that it can
change in an unstable fashion with each. The example in Figure
\ref{fig:LH} shows that the local homology group can depend strongly
on the choice of radius. From the same figure, we can also see that
it depends on the choice of center point; for example, if we fix a
small value of $r$ and move the center point $z$ gradually along the
line from its current location to the crossing point, the rank of
$LH_0(Y,z,r)$ jumps suddenly from two to four. The group $LH(Y,z,R)$
also clearly depends on the space $\Y$. For a stark example, replace
$\Y$ in Figure \ref{fig:LH} with a dense point sample $\U$. Then,
with probability one, the intersection $\U \cap S_R(z)$ will be
empty, and so $LH_0(U,z,R)$ will be zero.

Fortunately, the persistent version of local homology, which we now describe,
varies continuously with these input parameters, and so is much
more suitable to real-world data analysis.

\subsection{Persistent Local Homology}
\label{subsec:PLH}

\begin{figure}[h]
%\vspace{.3in}
\centerline{\fbox{\includegraphics[scale=0.3]{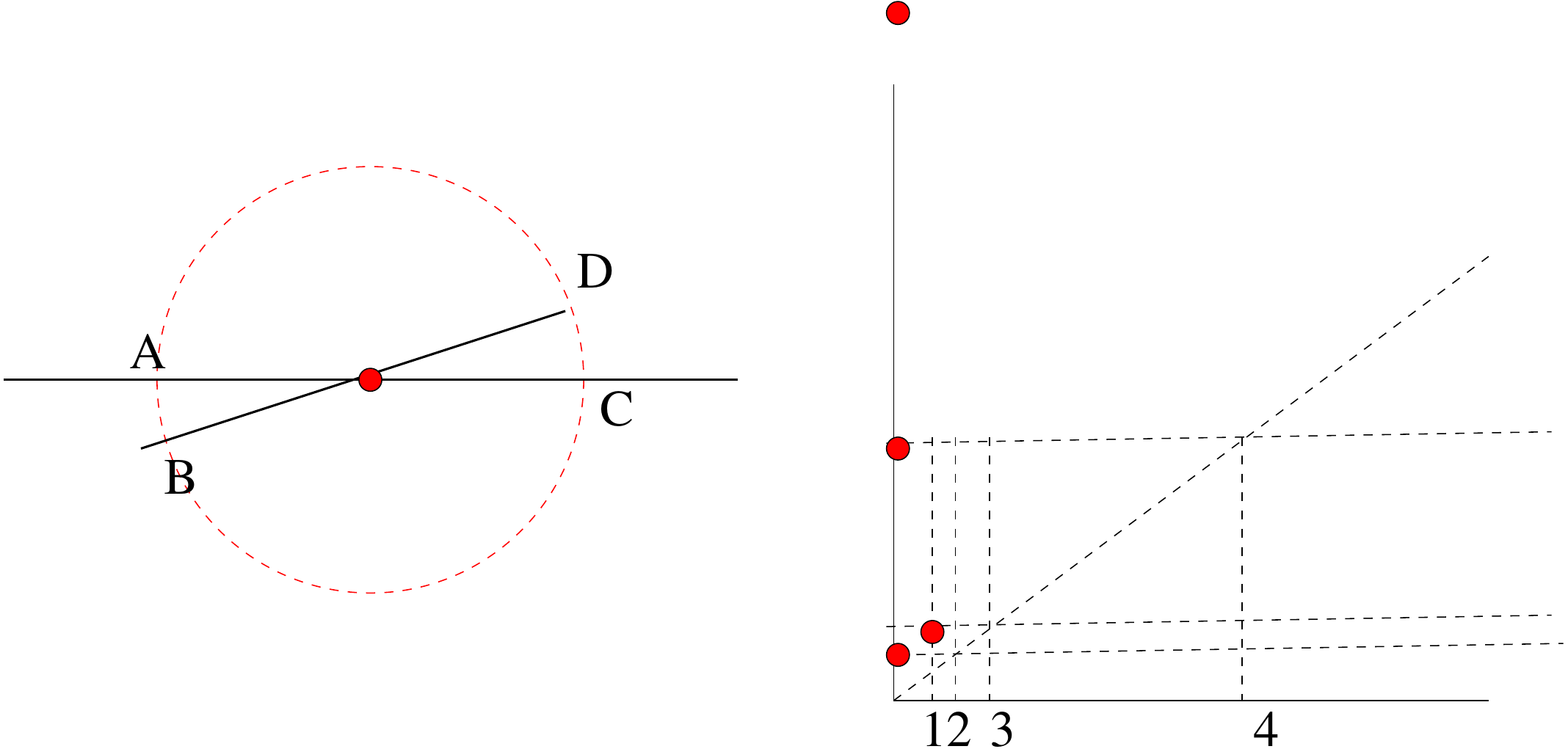}}}
%\vspace{.3in}
\caption{Left: $\Y$ is the intersecting line segments, $z$ is the marked point, and $S_R(z)$ is the dashed circle.
Right: the persistence diagram $PLH_0(Y,z,R)$; the dot on top has infinite death time, and the values of $\alpha_1$ through
$\alpha_4$ are indicated along the birth-axis.}
\label{fig:PLH}
\end{figure}

In what follows, we let $\Y$ be some compact space in $\R^D$.
For a working example, imagine that $\Y$ is the pair of crossing line
segments on the left side of Figure \ref{fig:PLH}, but it might
also be some point cloud $\U$ sampled from it.
For the moment, we fix a value of $R$ and a choice of $z$.
The basic idea behind PLH is that, instead of examining the connectivity of
$\Y \cap S_R(z)$,
we use a gradually thickening $\Y$ to filter $S_R(z)$ and watch the
homological information
change during this process.

More precisely, we let $d_Y : \R^D \to \R$ be the \emph{distance function} which
maps each point $x$ in $\R^D$
to the distance $d_Y(x)$ from its closest neighbor on $\Y$.
Abusing the notation above, we let $\Y_{\alpha}$ denote the threshold sets for
this function.
Note that $\Y_0 = \Y$, while $\Y_{\alpha}$ for positive $\alpha$ represents a
thickened version of $\Y$;
when $\Y$ is a point cloud, $\Y_{\alpha}$ is just the union of closed balls
of radius $\alpha$ around all of the points
in the cloud.

For each fixed value of $\alpha,$ we form the intersection $\Y_{\alpha} \cap S_R(z)$.
We then let $\alpha$ increase from zero to infinity and track the evolution of
the homology
groups $H_k(\Y_{\alpha} \cap S_R(z))$, calling the resulting persistence diagram
$PLH_k(\Y, z, R)$.
Note that this is just an alternative way of looking at the persistence
diagram $Dgm_k(d_Y|_{S_R(z)}).$

%\paragraph{Example.}

{\bf Example.} With $\Y, z,$ and $R$ as indicated in the caption for
the left side of Figure \ref{fig:PLH}, we work through the
computation of $PLH_0(Y,z,R)$. The original space $\Y \cap S_R(z)$
consists of three points, which we will call $A,B,C,$ working from
left to right. A small amount of thickening, say to $\alpha_1$,
produces a fourth component $D$. Almost immediately after, at
$\alpha_2$, the components $A$ and $B$ merge, followed quickly, at
$\alpha_3$ by the merging of $D$ and $C$. At that time, there are
two growing components, and they eventually merge when the entire
sphere is filled in at $\alpha_4$. In summary, we have the
birth-death pairs $\{(0, \infty), (0, \alpha_4), (0, \alpha_2),
(\alpha_1, \alpha_3)\},$ which leads to the diagram shown on the
right of the same figure.

%\paragraph{Stabilities.}

{\bf Stabilities.} As promised above, we now show that persistence
diagrams $PLH_k(\Y, z, R)$ are robust to small changes in any of
their three inputs. Recall that the Hausdorff distance $d_H(\Y,\Y')$
between two compact subsets is defined to be the minimum $\epsilon$
such that $\Y \subseteq \Y'_{\epsilon}$ and $\Y' \subseteq
\Y_{\epsilon}$,
\begin{theorem}
\label{thm:stab}
Let $\Y,\Y'$ be compact subsets of $\R^D$, let $z,z' \in \R^D$, and let $R, R' > 0$.
Set $\epsilon = d_H(\Y, \Y')$ and fix a non-negative integer $k$.
Then:
\begin{align*}
  W_{\infty}(PLH_k(\Y, z, R), PLH_k(\Y,z,R') &\leq |R - R'|, \\
  W_{\infty}(PLH_k(\Y, z, R), PLH_k(\Y,z',R) &\leq ||z - z'||, \\
  W_{\infty}(PLH_k(\Y, z, R), PLH_k(\Y',z,R) &\leq \epsilon. \\
  \end{align*}
\end{theorem}
\vskip -0.2in \proof For the first inequality, we may assume that
our fixed center point $z$ is at the origin in $\R^D$. We define two
functions $f,g: \R^D \to \R$ by $f(x) = d_Y(Rx)$ and $g(x) =
d_Y(R'x)$. Note that restricting $f$ and $g$ to the unit sphere $S$
in $\R^D$ is the same thing as restricting $d_Y$ to $S_R(z)$ and
$S_{R'}(z)$, respectively. So by Theorem \ref{thm:DST}, we have
\[W_{\infty}(PLH_k(\Y,z,R), PLH_k(\Y,z,R')) \leq ||f|_S - g|_S||_{\infty}.\]

To bound the right-hand-side, fix some $x \in S$, and let $y_0$ and $y_1$
be the closest points in $\Y$ to $Rx$ and $R'x$, respectively.
Then
\begin{align*}
 f(x) = ||y_0 - Rx|| &\leq ||y_1 - Rx||, \\
 g(x) = ||y_1 - R'x| &\leq ||y_0 - R'x||.
\end{align*}
Assume for the moment that $f(x) \geq g(x)$.
Then
\begin{align*}
 |f(x) - g(x)| &\leq |||y_1 - Rx|| - ||y_1 - R'x||| \\
 &\leq ||x||(|R - R'|) = |R - R'|.
\end{align*}
An identical argument takes care of the case when $g(x) > f(x),$
and taking a maximum over $S$ gives the claim.

A similar argument with the functions $h(x) = d_Y(z + Rx)$
and $j(x) = d_Y(z' + Rx)$ suffices for the second inequality.

Finally, it is easy to see that $||d_Y - d_{Y'}||_{\infty} \leq
d_H(\Y, \Y')$, for any pair of compact spaces. Since restricting the
domain of these two functions to $S_R(z)$ can only make the
$L_{\infty}$-distance smaller, a final application of Theorem
\ref{thm:DST} gives the third inequality. $\hfill \blacksquare$

These results can of course be combined, along with the triangle inequality,
to give a bound for what happens when all three inputs are changed at once.

%%%%%%%%%%%%%%%%%%%%%%%%%%%%%%%%%%%%%%%%%%%%%%%%%%%
%%%%%%%%%%%%%%%%%%%%%%%%%%%%%%%%%%%%%%%%%%%%%%%%%%%
\section{MLSA Features in Machine Learning}
\label{sec:results}

In what follows, we present two examples of synthetically generated point cloud datasets
 that were sampled from simple stratified spaces. We use them to investigate
 the role of MLPCA features and PLH features, separately and taken together,
 in SVM learning applications. We refer to the combination of MLPCA
 and PLH features as \emph{MLSA features}. We also include an example
 involving real data, namely, the LIDAR ground and vegetation datasets from \cite{BL},
 and demonstrate improved performance results when MLSA features,
  as opposed to MLPCA or PLH features on their own, are used to distinguish
  between the two classes.

%%%%%%%%%%%%%%%%%%%%%%%%%%%%%%%%%%%%%%%%%%%%%%%%%%%
\subsection{Data Preprocessing}
\label{subsec:preprocessing}

The traditional approach to data pre-processing consists of scaling
all the variables (features) to a common range of values (such as
[0,1] or [-1,1] etc.). Alternatively, for each of the features, its
empirical average can be subtracted from each data point, and the
result divided by the empirically calculated standard deviation:
that way, each coordinate of the data will have its mean equal to
zero with variance equal to one, as is the case for the standard normal
distribution $N(0,1)$. The justification of these pre-processing
methods is numerical: the precision of computational operations is
less prone to errors if numbers of similar magnitude are being used.

It is also important to consider robustness of the decision rules under
the conditions of data variability. Although there is no
comprehensive theory of treating such issues (some data
adaptation approaches are proposed in \cite{Sugiyama}), there are
some empirical observations (stemming from machine learning problems
in rather diverse application areas including computer networking,
medical diagnostics, computer vision etc.) that suggest that
deliberate pruning of available information can make the resulting
decision rules more robust and reliable, given the inevitable
variability between distributions of training and test datasets.

Specifically, we consider discretizing or binning of the already
scaled data (with their first moments matching $N(0,1)$) along each
of the available coordinates into several values that are centers of
equal-integral/equal-area segments of the standard normal density
function $N(0,1)$. In this paper, we consider discretization into 10
equal-integral bins, so that the boundaries $[b_i,b_{i+1}]$ of these bins
are such that the probability of an $N(0,1)$-distributed
random value belonging to any of them is $1/10$, i.e.,
$$\frac{1}{\sqrt{2\pi}}\int_{b_i}^{b_{i+1}}\exp\left(-\frac{x^2}{2}\right)\,dx=
\frac{1}{10}.$$ This condition is realized by the following bins
boundary values:
$$\begin{array}{l}-\infty, -1.2816,
-0.8416, -0.5244, -0.2533, 0, \\
+0.2533, +0.5244, +0.8416, +1.2816, +\infty \end{array}$$

Although discretization clearly reduces the information available in
the given datasets, the accuracy (error rate) of the classification
decision rule constructed on the discretized dataset is usually
comparable with the accuracy of the classification decision rule
constructed on the original dataset. Moreover, while the error rate is
about the same, the balance of sensitivity and specificity is
usually more stable for the discretized dataset. One can also argue
that discretization provides graceful handling of outliers without
losing the pertinent information and retaining the general direction
of the outlier. Finally, discretization appears to be more robust to
statistical deviations between training set and test set (the key
assumption of machine learning is that both these sets should have
the same distribution).

%%%%%%%%%%%%%%%%%%%%%%%%%%%%%%%%%%%%%%%%%%%%%%%%%%%
\subsection{Synthetic Dataset Examples}
\label{subsec:toy}

{\bf Datasets with Crossing Points.} We sampled $200$ points from
four different spaces, each consisting of unions of line segments
within a disk of radius $0.4$ centered at the origin. The four
spaces are: a ``$+$'' shape formed from portions of the $x$ and $y$
axes; an ``X'' shape formed by portions of the lines $y=-2x$ and
$y=3x$ near the origin; a ``Y'' shape made up of portions of the
lines $y=-x$ and $y=x$ above the $x$ axis together with part of the
negative $y$ axis; and a ``triple'' of line segments consisting of
the plus sign together with part of the line $y=x$. At each of the
points in the datasets, we performed MLPCA at three different radii
($0.1,$ $0.2,$ and $0.3$) and extracted the eigenvalues and
components of the corresponding eigenvectors, yielding $18$
MLPCA features per point. Furthermore, for each of the points at the
same three radii, we performed PLH, extracted the six most
persistent $0$-dimensional classes, and recorded as features the
persistences of these six classes. The reason for selecting six
classes is that at every point in each of the datasets, the
$0$-dimensional local homology groups are of rank at most three 
for the Y, at most four for the $+$ and the X, and at most six for the 
triple crossing. Thus, each of the feature vectors has length $36$.

In every instance, we trained a linear SVM classifier. We generated $50$
examples from each category of dataset for training and $15$ of each type for testing,
for a total of $10,000$ points for training and $3,000$ for testing for each
of the four categories.

To evaluate the performance of our features in the SVM
classification process, we computed the maximum error rate for Type
I (misclassify dataset A as dataset B) and Type II (misclassify
dataset B as dataset A) errors. We also recorded the sensitivity
($100\%$ minus the Type I error rate) and specificity ($100\%$ minus
the Type II error rate) as additional measures of accuracy.

The results of the six pairwise data experiments are reported in
Tables~\ref{table:plusXtable}-\ref{table:Ytripletable} in Section \ref{sec:ST}.
In every case, the combination of MLPCA and PLH features from MLSA led to the
lowest error rates, with PLH alone outperforming MLPCA alone in five
out of the six cases (the exception being the X vs. the triple crossing,
see Table~\ref{table:Xtripletable}). The discretization procedure did not
appear to have a strong impact on the results: binning improved MLSA
accuracy rates in three out of the six cases, and tied with
the no binning accuracy rates in one instance.

The lowest error rate, $0.03\%,$ was achieved for two different pairs:
the $+$ and the triple crossing, as well as the X and the Y. For the former,
the underlying $+$ shape is a subset of the triple crossing shape, but regardless of location within
each dataset, PLH detects the presence of more local homology classes (at least
at larger scales) in the case of the triple crossing than in the $+$ case. The same
is true for the case of the X and the Y, as well as for the pair consisting of the
Y and the triple, which also saw an error rate under $1\%$. For the case of the $+$
and the X, it is likely that the low MLSA error rates (again, under $1\%$)
were largely due to differences in birth and death times between their respective
local homology classes. 

The highest error rates (around $5-6\%$) occurred for the
pairs consisting of the $+$ and the Y, and the X and the triple crossing. However,
for the former, recall that both point cloud types contain points on the negative $y$ axis within a disk
of radius $0.4$. Since the largest radius in the MLPCA and PLH computations was $0.3$, 
such points on both the $+$ and the Y sufficiently far from the origin should indeed 
be indistinguishable from one another.
For the latter, the decrease in accuracy may be attributed to the fact that 
there are a number of points in both the X and the triple crossing point clouds
such that computing PLH at those points results in local homology groups of the 
same rank.

\begin{figure}[ht]
%\vspace{.3in}
\centerline{\fbox{
(a)  \includegraphics[scale=0.15]{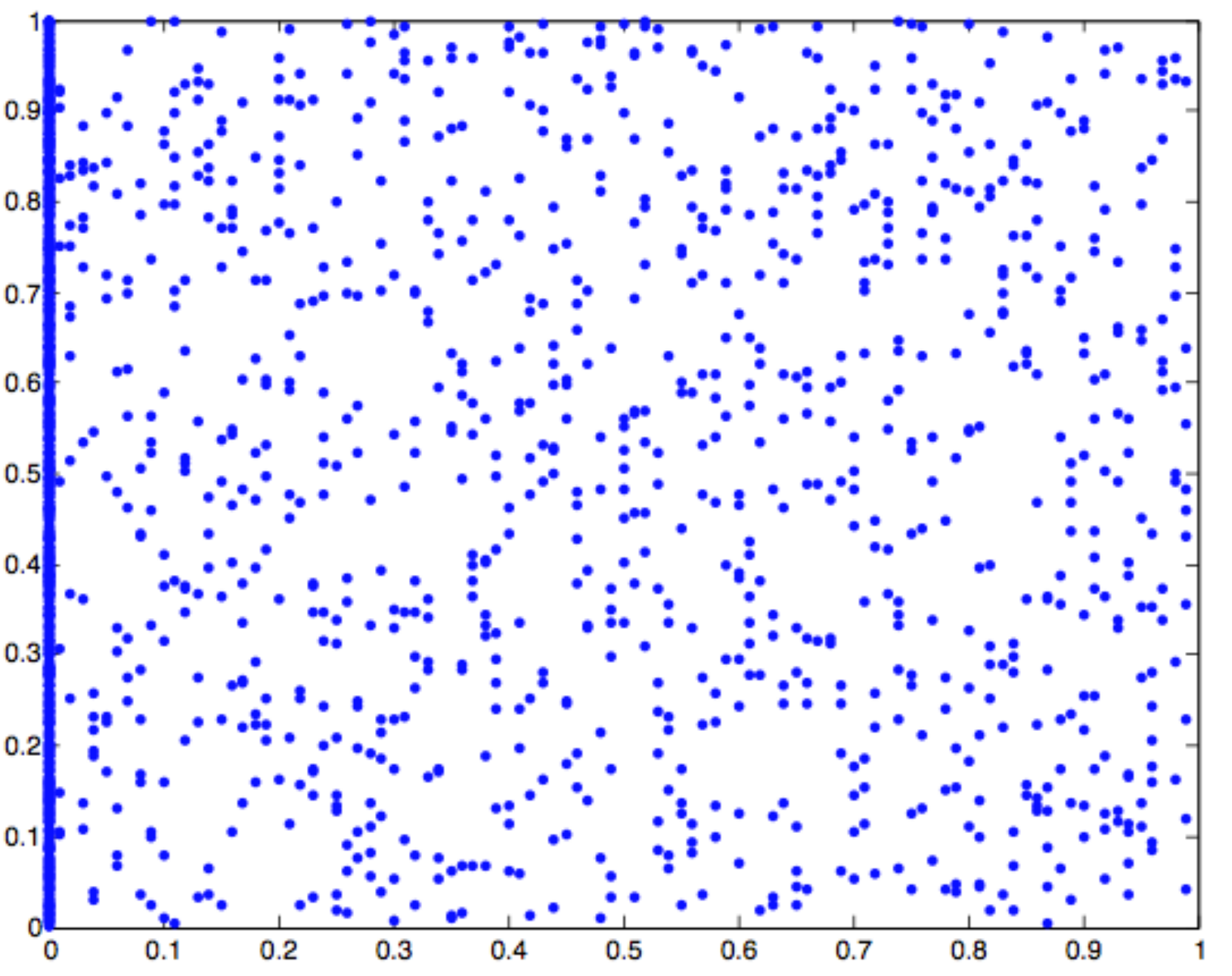}
 (b)    \includegraphics[scale=0.15]{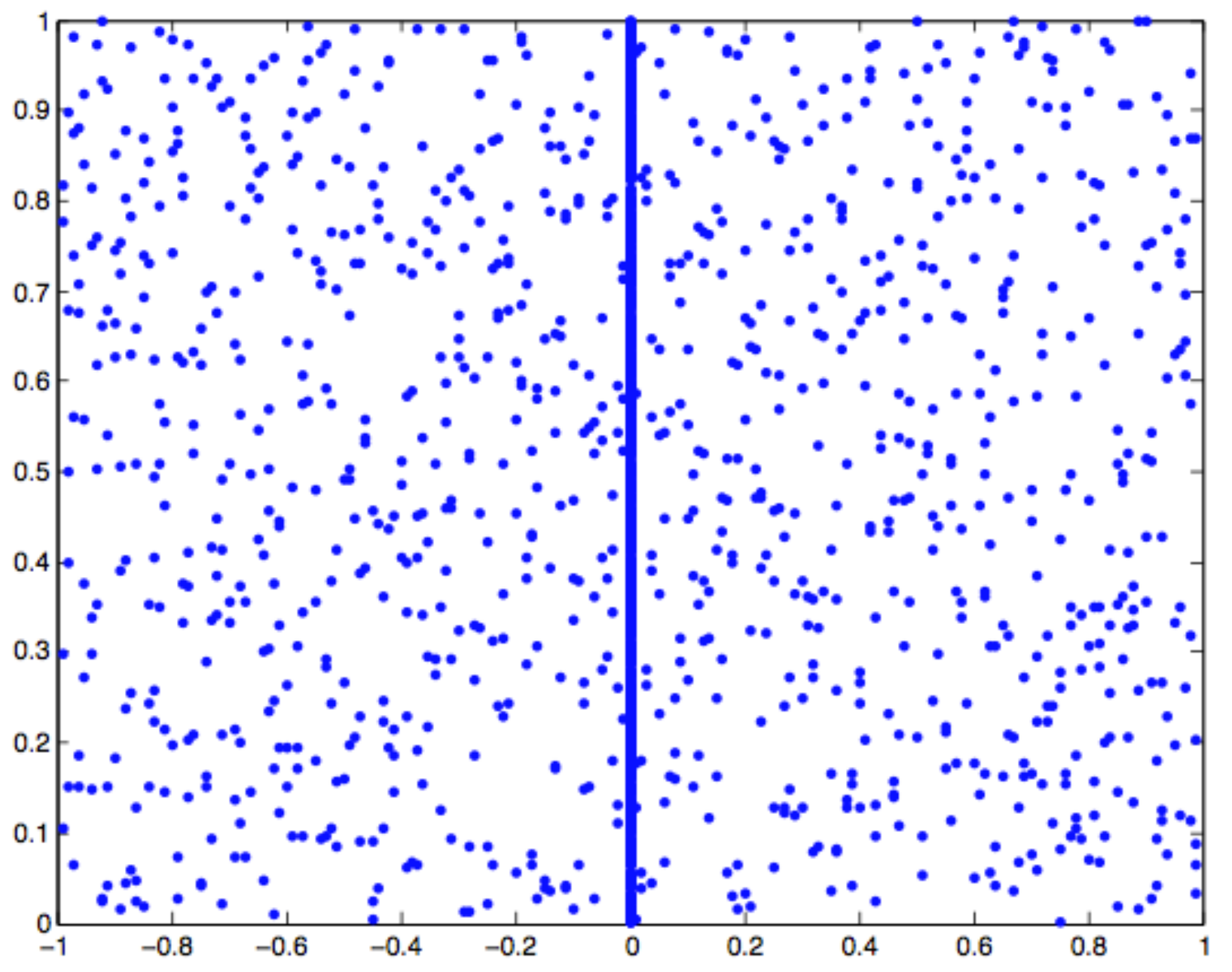}}}
      \caption{Densely sampled line segment with points on one side only vs. both sides.}
      \label{fig:sides}
%\vspace{.3in}
      \end{figure}

{\bf Densely Sampled Line Segments with Points on One Side vs. Both
Sides.} For our second synthetic dataset example, we obtained point clouds 
in two different ways: first, by sampling $200$ points from the line segment 
$x=0, 0 \leq y \leq 1$, along with $200$ points from the
unit square $[0,1]\times [0,1]$ (see Figure~\ref{fig:sides}(a)); second, 
by sampling $200$ points from the same line
segment as well as a total of $200$ points from the rectangle
$[-1,1] \times [0,1]$ ($100$ points on either side of the line segment,
see Figure~\ref{fig:sides}(b)). The goal of the machine learning experiment was to distinguish
points on the densely sampled line segments in the first case from points
on the corresponding line segments in the second case.

In both cases, for radii $0.2$ and $0.4$, 
we computed MLPCA features (two eigenvalues
and components of the two associated eigenvectors) and PLH features
(the persistence of the most persistent $1$-dimensional PLH class)
for a total of 14 features
at each of the $200$ points on the densely sampled line segments.
The reasoning behind our choice of PLH features is as follows.
When PLH is computed at points on
the densely sampled line segment in Figure~\ref{fig:sides}(b), it
detects a high-persistence $1$-dimensional class, whereas the $1$-dimensional
local homology is trivial in the case of the line segment with points on only one side. 
Note that $0$-dimensional PLH data should be the same in both cases.

As in our previous set of examples, we trained a linear SVM classifier 
with $10,000$ points from each category 
of dataset for training and $3,000$ for testing.  
Once again, the lowest error rate ($4.07\%$) was achieved when both
MLPCA and PLH features were utilized, with PLH features alone still vastly outperforming 
MLPCA features alone ($7.83\%$ vs. $33.83\%$); see
Table~\ref{table:sidestable}. In this example,
$10-$bin discretization led to poorer accuracy rates for both
MLSA features as well as PLH features alone than when no binning was employed. 
For MLPCA features alone, the results
were slightly better when binning was utilized ($31.4\%$ vs. $33.83\%$).

%%%%%%%%%%%%%%%%%%%%%%%%%%%%%%%%%%%%%%%%%%%%%%%%%%%

\subsection{LIDAR Image Dataset Results}
\label{subsec:LIDAR}

\begin{figure}
%\vspace{.3in}
\centerline{\fbox{\includegraphics[scale=0.25,angle=270]{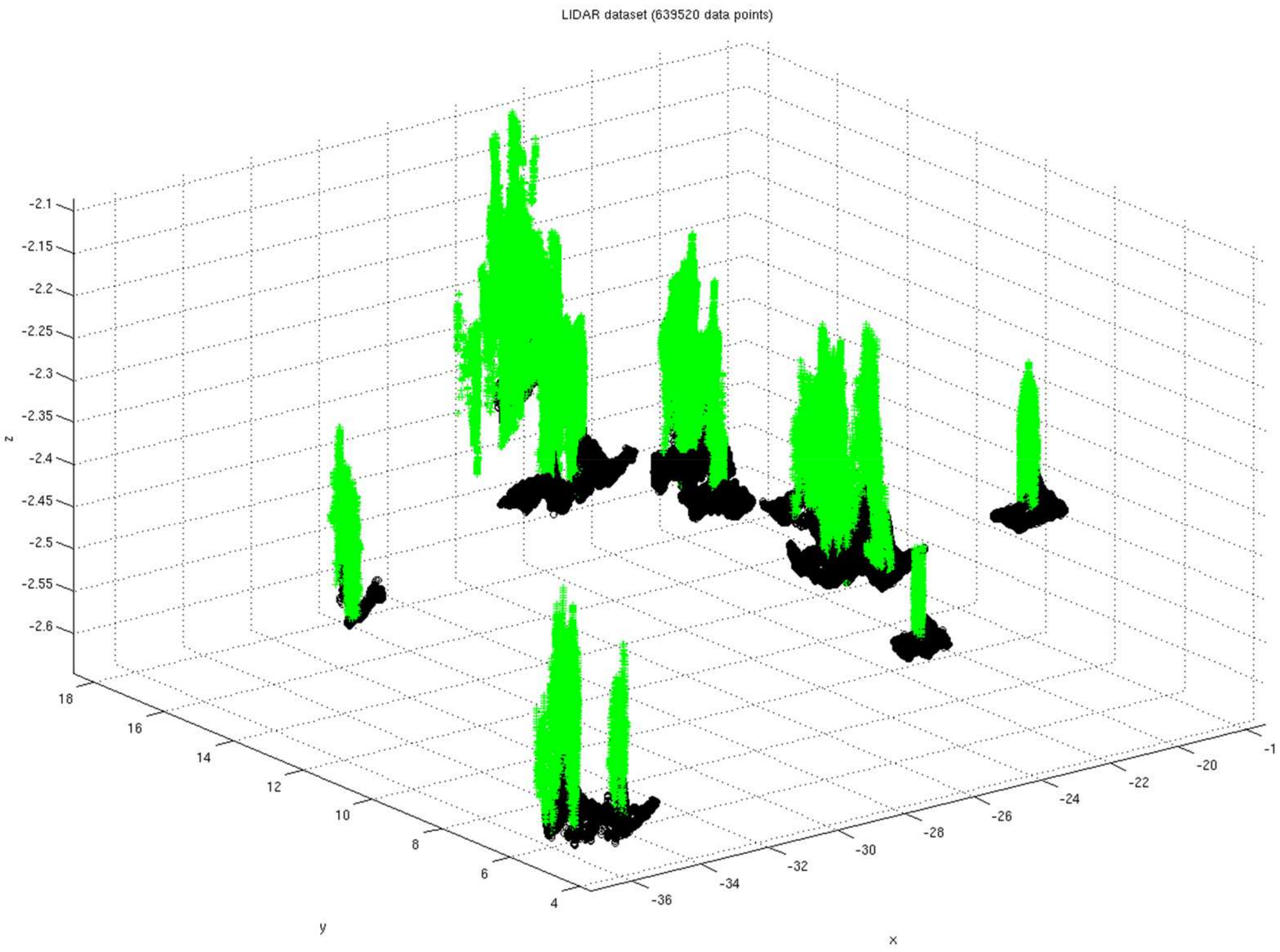}}}
%\vspace{.3in}
\caption{LIDAR dataset: green points correspond to ``vegetation,''
black points correspond to ``ground.''} \label{fig:LIDAR}
\end{figure}

In order to investigate the utility of PLH information for feature
construction in real data, we studied a dataset, introduced
in \cite{BL}, which consists of a total of $639,520$
three-dimensional points that were collected via LIDAR and labeled
as either ``ground'' or ``vegetation,'' with $10$ subsets of each
type (see Figure~\ref{fig:LIDAR}).

For proof-of-concept purposes, we randomly sampled $1000$ points from each
of the $10$ subsets in each category, for a total of $20,000$ points.
We split the data into training and testing groups in the same way as
the authors of \cite{BL} and \cite{BIMNS}; namely, we used the points
in data subsets $1, 2, 4, 6$, and $8$ for training, and subsets $3, 5, 7, 9$,
and $10$ for testing. At each of the two scales $2^{-3}$ and $2^{-4}$, 
we computed MLPCA features
(three eigenvalues and components of three eigenvectors)
and PLH features (persistences of the most persistent $0$-dimensional class
and the most persistent $1$-dimensional class)
at each of the $1000$ points. In addition, as in \cite{BIMNS}, we included the
coordinates of the points in the list of features. This yielded a total of
$31$ features.

For MLPCA features alone with discretization, the maximum of the
Type I (misclassify ground floor as vegetation) and Type II (misclassify
vegetation as ground floor) error rates was $4.95\%$. When the PLH
features were added to the MLPCA features,
the maximum error rate was reduced to $4.31\%$, a $15\%$ improvement.
See Table \ref{table:LIDARtable} for details. Note that taking only PLH
features led to greatly reduced levels of accuracy compared to taking
either MLSA features or MLPCA features alone. 
This is likely due to the low number of PLH features (namely, four)
compared to the size of the datasets. In all cases, discretization resulted
in an improvement of accuracy rates on the order of $15-20\%$.

%%%%%%%%%%%%%%%%%%%%%%%%%%%%%%%%%%%%%%%%%%%%%%%%%%%
%%%%%%%%%%%%%%%%%%%%%%%%%%%%%%%%%%%%%%%%%%%%%%%%%%%

%%%%%%%%%%%%%%%%%%%%%%%%%%%%%%%%%%%%%%%%%%%%%%%%%%%
%%%%%%%%%%%%%%%%%%%%%%%%%%%%%%%%%%%%%%%%%%%%%%%%%%%

%\pagebreak

%%%%%%%%%%%%%%%%%%%%%%%%%%%%%%%%%%%%%%%%%%%%%%%%%%%
%%%%%%%%%%%%%%%%%%%%%%%%%%%%%%%%%%%%%%%%%%%%%%%%%%%
%%%%%%%%%%%%%%%%%%%%%%%%%%%%%%%%%%%%%%%%%%%%%%%%%%%
%%%%%%%%%%%%%%%%%%%%%%%%%%%%%%%%%%%%%%%%%%%%%%%%%%%
%%%%%%%%%%%%%%%%%%%%%%%%%%%%%%%%%%%%%%%%%%%%%%%%%%%
%%%%%%%%%%%%%%%%%%%%%%%%%%%%%%%%%%%%%%%%%%%%%%%%%%%

%%%%%%%%%%%%%%%%%%%%%%%%%%%%%%%%%%%%%%%%%%%%%%%%%%%
%%%%%%%%%%%%%%%%%%%%%%%%%%%%%%%%%%%%%%%%%%%%%%%%%%%

\section{Discussion}
\label{sec:disc}

The above results are promising in that multi-scale local shape
analysis via a combination of MLPCA and PLH features consistently
led to improved classification decision results for both synthetic
and real datasets. 

The synthetic data experiments suggest future research to determine
if more sophisticated multi-scale local principal component
features could improve detection of local dimensions in singular
spaces. Moreover, there are other methods of turning persistence diagrams into
features
(for example, treating the diagram as a binned image \cite{Tracking2014}, or using ideas from
algebraic geometry \cite{Adcock2013}) 
that may be more advantageous in different settings.  

In this context, the advantages of $10-$bin discretization are less
pronounced. In fact, in synthetic datasets, the discretization process
led to poorer performance results, although it did improve the
classification quality for the real dataset. This is actually to be
expected, since the primary value of discretization is to make the
decision rule robust in the case when there are statistical
differences between training and test data.  With real life data, it
is often the case that there are such
  differences (e.g., one collects data for training, and at the time of testing
  the data, the selection mechanism may be slightly different from that for training,
  etc.). Thus, binning allows the decision rule to retain its robustness in such
   situations, as in the LIDAR case, where various patches of ground and vegetation
   differ from each other. If, however, there is a strong statistical match between
   training and test data (as typically happens in synthetic data,
   where the match is enforced by sampling the same distribution),
   then binning is useless: it throws away
   information for the sake of robustness, which is irrelevant in the case of a
    perfect match, causing performance to suffer.

We have identified several steps that we plan to undertake
in future work. Among them is a more advanced version of the MLPCA approach used
in this paper. It relies on features defined in terms of normalized
multi-scale constructs called Jones Beta numbers ~\cite{Jo90,
Le00, Le03, Paj02}. Our preliminary experiments with this approach
show improved classification rates and lower testing error rates for
a number of cases in which these are the only features utilized. 
We plan to continue with additional experiments and investigate what happens when the
new features are combined with PLH features.

\section{Summary Tables}
\label{sec:ST}

The following tables summarize the results of the synthetic dataset experiments (Tables~\ref{table:plusXtable}-\ref{table:sidestable}) and the LIDAR dataset experiment (Table~\ref{table:LIDARtable}). A value of 10 for ``Bins'' indicates the utilization of 10-bin discretization.

    \begin{table}[!h]
        \caption{$+$ vs. X}
    \begin{tabular}{|c|c|c|c|c|}
    \hline
Features&   Bins&Sens.&Spec.& Max Errors \\
                \hline
PLH &No&    99.00\% &98.63\%    &1.37\%\\
\hline
MLPCA&  No& 87.90\%&   95.67\%&    12.10\% \\
\hline
MLSA&    No& 99.37\%&    99.30\% &0.70\%\\
\hline
PLH &10 & 89.13\% & 94.67\%    &10.87\%    \\
\hline
MLPCA&  10  & 84.23\% &93.60\%    &15.77\%\\
\hline
MLSA &10& 98.23\% &98.63\%    &1.77\%\\
                            \hline

    \end{tabular}
    \label{table:plusXtable}

\end{table}

    \begin{table}[!h]
        \caption{$+$ vs. Y}
    \begin{tabular}{|c|c|c|c|c|}

    \hline
Features&   Bins&Sens.&Spec.& Max Errors \\
                \hline
PLH &No&  89.23\%    &92.93\%    &10.77\%\\
\hline
MLPCA&  No& 87.83\% &   94.23\% &   12.17\% \\
\hline
MLSA&    No& 93.90\% &   97.50\% &   6.10\%\\
\hline
PLH &10 & 85.87\% & 97.23\% &   14.13\%     \\
\hline
MLPCA&  10  & 82.27\% &   91.33\% &   17.73\%\\
\hline
MLSA &10& 93.93\% &   98.87\% &   6.07\%\\
                            \hline
    \end{tabular}
      \label{table:plusYtable}
\end{table}

    \begin{table}[!h]
        \caption{$+$ vs. Triple}
    \begin{tabular}{|c|c|c|c|c|}

    \hline
Features&   Bins&Sens.&Spec.& Max Errors \\
                \hline
PLH &No&   99.73\% &   99.43\% &   0.57\%\\
\hline
MLPCA&  No& 86.67\% &   96.80\% &   13.33\% \\
\hline
MLSA& No & 99.83\% &   99.93\% &   0.17\%  \\
\hline
PLH &10 & 99.97\% & 99.77\% &   0.23\% \\
\hline
MLPCA&  10  & 86.43\% &   95.27\% &   13.57\%\\
\hline
MLSA &10& 99.97\% &   99.97\% &   0.03\%  \\
                            \hline

    \end{tabular}
      \label{table:plustripletable}
\end{table}

    \begin{table}[!h]
        \caption{X vs. Y}
    \begin{tabular}{|c|c|c|c|c|}

    \hline
Features&   Bins&Sens.&Spec.& Max Errors \\
                \hline
PLH &No&  95.20\% &   98.93\% &   4.80\%\\
\hline
MLPCA&  No& 79.17\% &   93.77\% &   20.83\% \\
\hline
MLSA&    No& 99.97\% &   100.00\% &  0.03\%  \\
\hline
PLH &10 & 94.67\% &   97.77\% &   5.33\%  \\
\hline
MLPCA&  10  & 81.50\% &   94.80\% &   18.50\%\\
\hline
MLSA &10&  99.97\% &   99.97\% &   0.03\% \\
                            \hline

    \end{tabular}
      \label{table:XYtable}
\end{table}

    \begin{table}[!h]
        \caption{X vs. Triple}
    \begin{tabular}{|c|c|c|c|c|}

    \hline
Features&   Bins&Sens.&Spec.& Max Errors \\
                \hline
PLH &No&   89.13\% &   92.03\% &   10.87\%\\
\hline
MLPCA&  No&  90.00\% &   92.53\% &   10.00\% \\
\hline
MLSA&    No& 94.20\% &   98.30\% &   5.80\%  \\
\hline
PLH &10 & 89.50\% &   93.90\% &   10.50\% \\
\hline
MLPCA&  10  & 92.60\% &   93.30\% &   7.40\% \\
\hline
MLSA &10&  95.00\% &   98.73\% &   5.00\%\\
                            \hline

    \end{tabular}
      \label{table:Xtripletable}
\end{table}

    \begin{table}[!h]
        \caption{Y vs. Triple}
    \begin{tabular}{|c|c|c|c|c|}

    \hline
Features&   Bins&Sens.&Spec.& Max Errors \\
                \hline
PLH &No& 98.17\% &   100.00\% &  1.83\%\\
\hline
MLPCA&  No& 92.70\% &   91.37\% &   8.63\%  \\
\hline
MLSA&    No& 99.23\% &   100.00\% &  0.77\%\\
\hline
PLH &10 & 97.47\% &   100.00\% &  2.53\%  \\
\hline
MLPCA&  10  & 95.77\% &   87.57\% &   12.43\%\\
\hline
MLSA &10& 99.13\% &   100.00\% &  0.87\%\\
                            \hline

    \end{tabular}
      \label{table:Ytripletable}
\end{table}

    \begin{table}[!h]
        \caption{One Side vs. Both Sides}
    \begin{tabular}{|c|c|c|c|c|}

    \hline
Features&   Bins&Sens.&Spec.& Max Errors \\
                \hline
PLH &No&  92.17\%&   99.73\%&    7.83\%\\
\hline
MLPCA&  No& 66.17\%&    86.03\%&    33.83\% \\
\hline
MLSA&    No& 95.93\%&    98.73\% &4.07\%\\
\hline
PLH &10 & 91.67\%&    99.73\%&    8.33\%  \\
\hline
MLPCA&  10  & 68.60\%&    83.97\%&    31.40\%\\
\hline
MLSA &10& 95.13\%&    98.73\%&    4.87\%\\
                            \hline
    \end{tabular}
      \label{table:sidestable}
\end{table}

\begin{table}[!h]
        \caption{LIDAR}
    \begin{tabular}{|c|c|c|c|c|}

    \hline
Features&   Bins&Sens.&Spec.& Max Errors \\
                \hline
PLH &No&    92.80\% &  74.04\% &  25.96\%\\
\hline
MLPCA&  No& 94.28\% &  98.86\% &  5.72\%  \\
\hline
MLSA&    No& 94.90\% &  98.56\% &  5.10\% \\
\hline
PLH &10 & 92.98\% &  78.04\% &  21.96\% \\
\hline
MLPCA&  10  & 95.05\% &  99.06\% &  4.95\%\\
\hline
MLSA &10&  95.69\% &  99.14\% &  4.31\% \\
                            \hline

    \end{tabular}
      \label{table:LIDARtable}
\end{table}

\newpage
\clearpage

\bibliography{PLHAndMLPCA}{}

\begin{thebibliography}{10}

\bibitem{Adcock2013}
A.~Adcock, E.~Carlsson, and G.~Carlsson.
\newblock The ring of algebraic functions on persistence bar codes.
\newblock {\em ArXiv e-prints}, 2013.

\bibitem{AhmedFasy2014}
Mahmuda Ahmed, Brittany Fasy, and Carola Wenk.
\newblock Local persistent homology based distance between maps.
\newblock In {\em Proc. ACM SIGSPATIAL GIS}, 2014.
\newblock to appear.

\bibitem{BIMNS2}
D.~Bassu, R.~Izmailov, A~McIntosh, L.~Ness, and D.~Shallcross.
\newblock Application of multi-scale singular vector decomposition to vessel
  classification in overhead satellite imagery.
\newblock {\em submitted}.

\bibitem{BIMNS}
D.~Bassu, R.~Izmailov, A~McIntosh, L.~Ness, and D.~Shallcross.
\newblock Centralized multi-scale singular value decomposition for feature
  construction in {L}{I}{D}{A}{R} image classification problems.
\newblock In {\em Applied Imagery Pattern Recognition Workshop (AIPR), 2012
  IEEE}, pages 1--6, Oct 2012.

\bibitem{BCE07}
P.~Bendich, D.~Cohen-Steiner, H.~Edelsbrunner, J.~Harer, and D.~Morozov.
\newblock Inferring local homology from sampled stratified spaces.
\newblock In {\em Proceedings 48th Annual IEEE Symposium on Foundations of
  Computer Science}, pages 536--546, 2007.

\bibitem{bendich2012stratlearn}
P.~Bendich, B.~Wang, and S.~Mukherjee.
\newblock Local homology transfer and stratification learning.
\newblock In {\em Proceedings of the Twenty-Third Annual ACM-SIAM Symposium on
  Discrete Algorithms}, pages 1355--1370. SIAM, 2012.

\bibitem{Tracking2014}
Paul Bendich, Sang Chin, Jesse Clarke, John deSena, John Harer, Elizabeth
  Munch, Andrew Newman, David Porter, David Rouse, Nate Strawn, and Adam
  Watkins.
\newblock Topological and statistical behavior classifiers for tracking
  applications.
\newblock 2014.
\newblock arXiv:1406.0214.

\bibitem{Bow}
S.-T. Bow.
\newblock {\em Pattern Recognition and Image Preprocessing}.
\newblock New York: Marcel Dekker, 2002.

\bibitem{Breiman}
L.~Breiman.
\newblock Random forests.
\newblock {\em Machine Learning}, {\bf 45}:5--32, 2001.

\bibitem{BL}
N.~Brodu and D.~Lague.
\newblock 3{D} terrestial {L}{I}{D}{A}{R} data classification of complex
  natural scenes using a multi-scale dimensionality criterion: Applications in
  geomorphology.
\newblock {\em ISPRS Journal of Photogrammetry and Remote Sensing}, {\bf
  68}:121--134, 2012.

\bibitem{Chazal2009b}
F.~Chazal, D.~Cohen-Steiner, M.~Glisse, L.~Guibas, and S.~Oudot.
\newblock Proximity of persistence modules and their diagrams.
\newblock In {\em Proceedings of the 25th annual symposium on Computational
  geometry}, SCG '09, pages 237--246, New York, NY, USA, 2009. ACM.

\bibitem{CohenSteiner2007}
D.~Cohen-Steiner, H.~Edelsbrunner, and J.~Harer.
\newblock Stability of persistence diagrams.
\newblock {\em Discrete Comput. Geom.}, 37(1):103--120, January 2007.

\bibitem{Cohen-Steiner2010}
D.~Cohen-Steiner, H.~Edelsbrunner, J.~Harer, and Y.~Mileyko.
\newblock Lipschitz functions have ${L}_p$-stable persistence.
\newblock {\em Found. Comput. Math.}, 10(2):127--139, February 2010.

\bibitem{CoifmanLafon}
R.~Coifman and S.~Lafon.
\newblock Diffusion maps.
\newblock {\em Applied and Computational Harmonic Analysis}, {\bf 21}:5--30,
  2006.

\bibitem{DFW14}
T.~K. {Dey}, F.~{Fan}, and Y.~{Wang}.
\newblock {Dimension Detection with Local Homology}.
\newblock {\em arXiv e-prints}, May 2014.

\bibitem{Domingos}
P.~Domingos.
\newblock A few useful things to know about machine learning.
\newblock {\em Communications of the ACM}, {\bf 55}:78--87, 2012.

\bibitem{Edelsbrunner2010}
H.~Edelsbrunner and J.~Harer.
\newblock {\em Computational Topology: An Introduction}.
\newblock American Mathematical Society, 2010.

\bibitem{GuyonElisseeff}
I.~Guyon and A.~Elisseeff.
\newblock An introduction to variable and feature selection.
\newblock {\em Communications of the ACM}, {\bf 3}:1157--1182, 2003.

\bibitem{Jones08}
P.~Jones, M.~Maggioni, and R.~Schul.
\newblock Manifold parameterizations by eigenfunctions of the {L}aplacian and
  heat kernels.
\newblock {\em Proceedings of the National Academy of Sciences}, {\bf
  105}:1803--1808, 2008.

\bibitem{Jo90}
P.W. Jones.
\newblock Rectifiable sets and the traveling salesman problem.
\newblock {\em Invent. Math.}, {\bf 102}:1--15, 1990.

\bibitem{Le00}
J.C. L\'{e}ger.
\newblock Menger curvature and rectifiability.
\newblock {\em Ann. of Math.}, {\bf 149}:831--869, 1999.

\bibitem{Le03}
G.~Lerman.
\newblock Quantifying curvelike structures of measures by using {$L_2$} {J}ones
  quantities.
\newblock {\em Communications on Pure and Applied Mathematics},
  56(9):1294--1365, 2003.

\bibitem{LiuMotoda}
H.~Liu and H.~Motoda.
\newblock {\em Computational Methods of Feature Selection}.
\newblock Chapman \& Hall/CRC, 2008.

\bibitem{Munkres2}
J.~R. Munkres.
\newblock {\em Elements of Algebraic Topology}.
\newblock Addison Wesley, 1993.

\bibitem{Paj02}
H.~Pajot.
\newblock {\em Analytic capacity, rectifiability, Menger curvature and the
  Cauchy integral}.
\newblock Number 1799 in Lecture Notes in Mathematics. Springer, 2002.

\bibitem{Scholkopf}
B.~Sch\"olkopf, A.~Smola, and K.~M\"uller.
\newblock Nonlinear component analysis as a kernel eigenvalue problem.
\newblock {\em Neural Computation}, {\bf 10}:1299--1319, 1998.

\bibitem{Sonka}
M.~Sonka, V.~Hlavac, and R.~Boyle.
\newblock {\em Image Processing, Analysis and Machine Vision}.
\newblock Toronto, Ontario: Thomson Learning, 2007.

\bibitem{Sayan2014grassman}
B.~{St.~Thomas}, L.~{Lin}, L.-H. {Lim}, and S.~{Mukherjee}.
\newblock {Learning Subspaces of Different Dimension}.
\newblock {\em arXiv e-prints}, April 2014.

\bibitem{Sugiyama}
M.~Sugiyama, T.~Suzuki, and T.~Kanamori.
\newblock {\em Density Ratio Estimation in Machine Learning}.
\newblock Cambridge, MA: Cambridge University Press, 2012.

\bibitem{Torkkola}
K.~Torkkola.
\newblock Feature extraction by non-parametric mutual information maximization.
\newblock {\em Journal of Machine Learning Research}, {\bf 3}:1415--1438, 2003.

\end{thebibliography}
\bibliographystyle{plain}

\end{document}